\documentclass[%
 aip,
 apl,
 amsmath,
 amssymb,
 reprint,
]{revtex4-1}

\usepackage{graphicx}
\usepackage{dcolumn}
\usepackage{bm}
\usepackage[version=4]{mhchem} 
\usepackage{siunitx} 
\usepackage[hidelinks]{hyperref}

\usepackage{xcolor}
\usepackage{booktabs}
\usepackage{amsmath}
\usepackage{array}

\newcommand{\cro}{\ce{Ca2RuO4} }

\newcommand{\ngo}{\ce{NdGaO3} }
\newcommand{\lsat}{\ce{(LaAlO3)_{0.3}(Sr2AlTaO6)_{0.7}}}
\newcommand{\lsao}{\ce{LaSrAlO4} }
\newcommand{\ncao}{\ce{NdCaAlO4} }
\newcommand{\lao}{\ce{LaAlO3} }
\newcommand{\croinf}{\ce{CaRuO3} }

\newcommand{\sroinf}{\ce{SrRuO3} }

\newcommand{\alo}{\ce{Al2O3} }

\DeclareSIUnit\torr{torr}
\DeclareSIUnit\oersted{Oe}
\DeclareSIUnit\emu{emu}
\DeclareSIUnit\mubohr{$\mu_{\text{B}}$}

\begin{document}

\title{Tailoring the electronic properties of \cro via epitaxial strain}

\author{C. Dietl}
\affiliation{Max Planck Institute for Solid State Research, Heisenbergstrasse 1, 70569 Stuttgart, Germany}
\author{S. K. Sinha}
\affiliation{Max Planck Institute for Solid State Research, Heisenbergstrasse 1, 70569 Stuttgart, Germany}
\author{G. Christiani}
\affiliation{Max Planck Institute for Solid State Research, Heisenbergstrasse 1, 70569 Stuttgart, Germany}
\author{Y. Khaydukov}
\affiliation{Max Planck Institute for Solid State Research, Heisenbergstrasse 1, 70569 Stuttgart, Germany}
\affiliation{Max Planck Society Outstation at the Heinz Maier-Leibnitz Zentrum (MLZ), 85748 Garching, Germany}
\author{T. Keller}
\affiliation{Max Planck Institute for Solid State Research, Heisenbergstrasse 1, 70569 Stuttgart, Germany}
\affiliation{Max Planck Society Outstation at the Heinz Maier-Leibnitz Zentrum (MLZ), 85748 Garching, Germany}
\author{D. Putzky}
\affiliation{Max Planck Institute for Solid State Research, Heisenbergstrasse 1, 70569 Stuttgart, Germany}
\author{S. Ibrahimkutty}
\affiliation{Max Planck Institute for Solid State Research, Heisenbergstrasse 1, 70569 Stuttgart, Germany}
\author{P. Wochner}
\affiliation{Max Planck Institute for Solid State Research, Heisenbergstrasse 1, 70569 Stuttgart, Germany}
\author{G. Logvenov}
\affiliation{Max Planck Institute for Solid State Research, Heisenbergstrasse 1, 70569 Stuttgart, Germany}
\author{P. A. van Aken}
\affiliation{Max Planck Institute for Solid State Research, Heisenbergstrasse 1, 70569 Stuttgart, Germany}
\author{B. J. Kim}
\affiliation{Max Planck Institute for Solid State Research, Heisenbergstrasse 1, 70569 Stuttgart, Germany}
\affiliation{Department of Physics, Pohang University of Science and Technology, Pohang 790-784, Republic of Korea} 
\affiliation{Center for Artificial Low Dimensional Electronic Systems, Institute for Basic Science (IBS), 77 Cheongam-Ro, Pohang 790-784, Republic of Korea}
\author{B. Keimer}
\email{b.keimer@fkf.mpg.de}
\affiliation{Max Planck Institute for Solid State Research, Heisenbergstrasse 1, 70569 Stuttgart, Germany}
\date{\today}

\begin{abstract}
We report the synthesis of \cro thin films on \ncao (110), \lao (100) and \lsao (001) substrates and show that epitaxial strain induces a transition from the Mott-insulating phase of bulk \cro into a metallic phase. Magnetometry and spin-polarized neutron reflectometry reveal a low-temperature, small-moment ferromagnetic state in metallic \cro films.
\end{abstract}

\maketitle

Ruthenium oxides offer a rich platform for fundamental research, but also exhibit several properties of interest for technological application. A prime example is \sroinf which has been studied extensively for its itinerant ferromagnetism (FM) and peculiar transport properties, but also has been found to be an excellent electrode material for oxide electronics.\cite{Koster2012} Its close analogue \cro (CRO) hosts copious phenomena such as a metal-to-insulator transition (MIT), multiple magnetically ordered phases, orbital ordering phenomena as well as magnetic excitations analogous to the Higgs boson.\cite{Steffens2005,Braden1998,Zegkinoglou2005,Alexander1999,Souliou2017,Jain2017} Besides offering new insights into fundamental physics, the large electroresistive effect, MIT close to room temperature and high impedance of CRO might be useful for electronic devices.\cite{Fobes2012,Tanabe2017,Nakamura2013} Most research on CRO was conducted on single crystals, which are unsuited for integration into electronic circuits. Here we report the synthesis of CRO thin films which exhibit a higher potential for applications, and for the controlled modification of the lattice structure and electronic properties via epitaxial strain.

Rotation, tilt and flattening of the \ce{RuO_6} octahedra in the perovskite structure of CRO have been identified as key parameters determining the electronic ground state of CRO.\cite{Fang2001,Steffens2011} Hydrostatic and uniaxial pressure devices have been employed to tune these parameters and drive bulk CRO into a metallic ground state with a low-temperature ferromagnetic (FM) phase.\cite{Taniguchi2013,Alireza2010,Steffens2005,Nakamura2002} Motivated by the strong modification of the electronic and magnetic properties by relatively low pressure, we have studied the effects of biaxial epitaxial strain on CRO thin films.

The synthesis of CRO films was previously reported, but only growth of films on \lao(100) (LAO) substrates was achieved.\cite{Miao2012a,Wang2004} We show that growth on \lsao(001) (LSAO) and \ncao(110) (NCAO) substrates is also feasible, and discuss their structural and electronic properties.

CRO films of thickness 40, 50 and \SI{60}{\nano \meter} were grown on LSAO, LAO and NCAO, respectively, using pulsed laser deposition (PLD) from polycrystalline targets.

\begin{figure}[b]
	\centering
    \vspace*{-.2 cm}
		\includegraphics[width=\columnwidth]{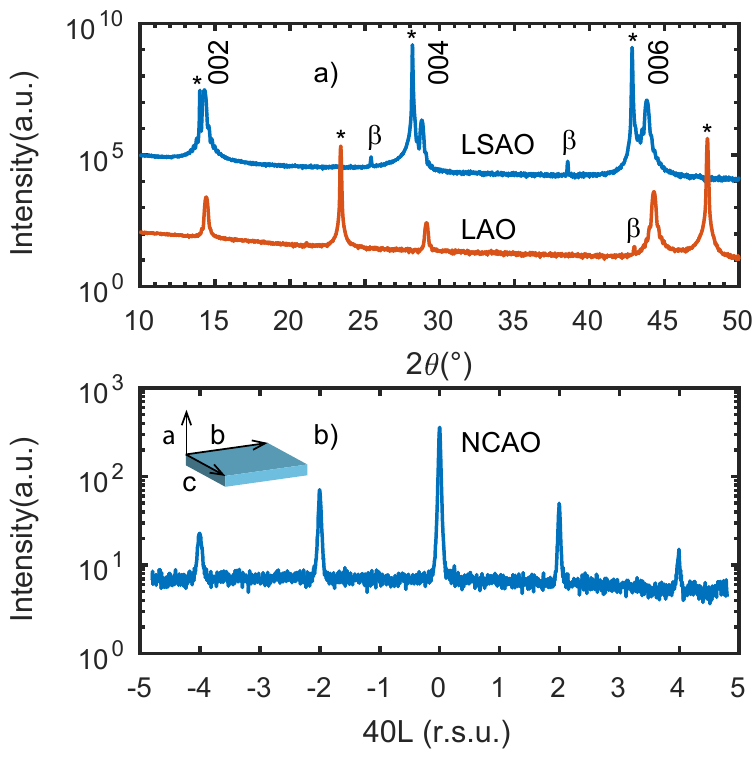}
	\caption{a) Specular $\theta-2\theta$ scans of CRO on LSAO and CRO on LAO b) Scan along the 40L direction of CRO on NCAO using the notation of the orthorhombic unit cell of CRO.  ($\beta$) denote peaks due to Cu-K-$\beta$ radiation, (*) indicate substrate reflections. The inset shows the orientation of the CRO unit cell grown on NCAO.}
		\label{fig:XRD_TTH}
\end{figure}

The targets were synthesized via a solid-state synthesis route by mixing \SI{99.999}{\percent} \ce{CaCO_3} and \SI{99.9}{\percent} \ce{RuO_2} in a 2:1.2 ratio to include Ru-excess for compensating the volatility of \ce{RuO_x} species during the ablation.
Both powders were weighed hot due to their hygroscopy. After a calcination step at \SI{900}{\celsius} for \SI{24}{\hour} in air, the mixture was pressed uniaxially in a \SI{30}{\milli\metre} diameter mold at \SI{62}{\mega\pascal}. The pellet was then pressed isostatically at \SI{640}{\mega\pascal} for further densification. Sintering was conducted at \SI{1370}{\celsius} for \SI{24}{\hour} in \SI{1}{\percent} \ce{O2} in \ce{Ar} atmosphere using non-compressed powder of the same composition as bedding powder, since CRO powders form an eutecticum with \alo crucibles.\cite{Nakatsuji2001} The density of the resulting pellet was estimated to be $\approx \SI{3.2}{\gram \per \centi \meter \cubed}$.

\begin{figure}[t]
	\centering
		\includegraphics[width=0.8\columnwidth]{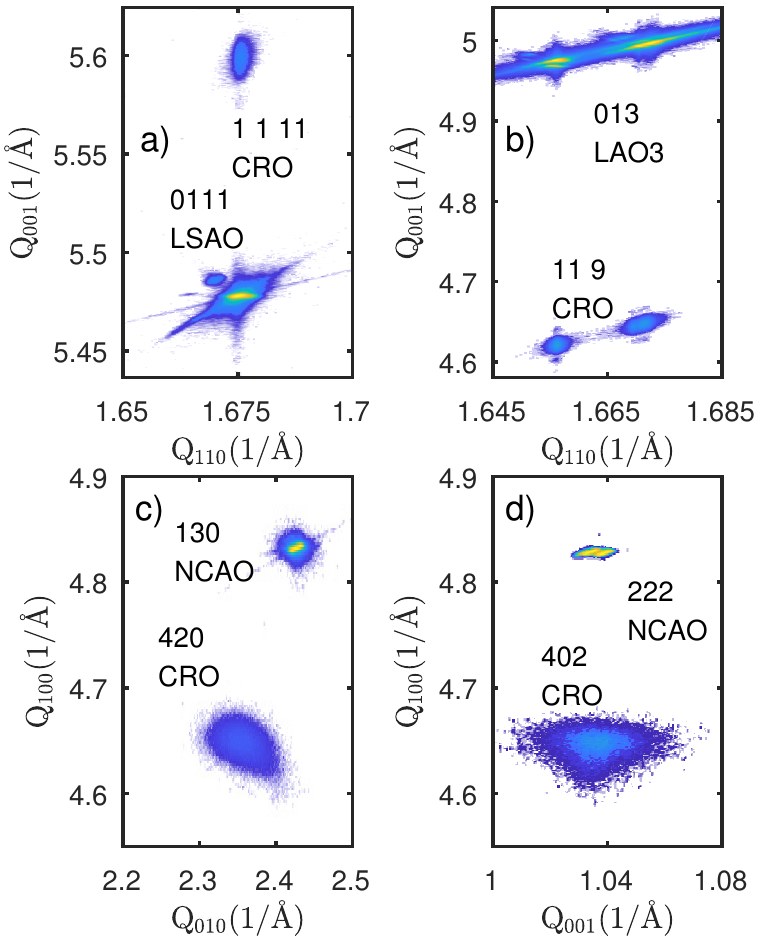}
	\caption{a) and b) Reciprocal Space Maps (RSM) showing the fully strained state of the films on LSAO and LAO. The double peak structure of LAO and consequently CRO arises from the inevitable twinning of the substrate\cite{Bueble1998} c) RSM of the film on NCAO showing the partially strained nature of the in-plane b-axis and d) the fully strained state along the in-plane c-axis. The axes refer to orthorhombic unit cell of CRO.}
		\label{fig:XRD_RSM}
            \vspace*{-.5 cm}
\end{figure}

The films on LAO and LSAO were grown at \SI{900}{\celsius}, while the film on NCAO was grown at \SI{860}{\celsius}. As a growth atmosphere, an oxygen partial pressure of  $p_{\text{O2}}=\SI{60}{\milli\torr}$ was employed. Temperature control was achieved via a pyrometer directed at the substrate using emissivities of $\epsilon_{\text{LAO}}=0.93$, $\epsilon_{\text{LSAO}}=0.92$ and $\epsilon_{\text{NCAO}}=0.92$. The substrate was glued to a high-purity tantalum piece with platinum paste to ensure good thermal contact. The Ta piece was then heated with an IR-laser to bring the substrate to growth temperature.  The ablation was conducted via an UV-Excimer-laser operating at \SI{248}{\nano\meter}, which was pulsed with \SI{11}{\hertz} using a fluence of \SI{\approx 2}{\joule\per\centi\meter\squared}. It was found to be essential to maintain good temperature homogeneity and reproducibility along the substrate to avoid the formation of the \croinf phase, which is favored at lower temperatures. Growths at \SI{950}{\celsius} necessitated high repetition rates $\geq$\SI{15}{\hertz} and only allowed us to grow films of \SI{\approx 10}{\nano\meter} thickness.  Repolishing the target after every run was deemed necessary to ensure a good crystalline quality, likely due to decomposition of the target stoichiometry under the laser ablation. All samples were rapidly quenched to room temperature by switching off the laser heater and injecting helium gas after the deposition while keeping the oxygen partial pressure to avoid decomposition. Attempts to grow on LSAT\footnote{\lsat} (100), \ngo (110) and \ngo (001) substrates were unsuccessful.

\begin{figure}[t]
	\centering
		\includegraphics[width=\columnwidth]{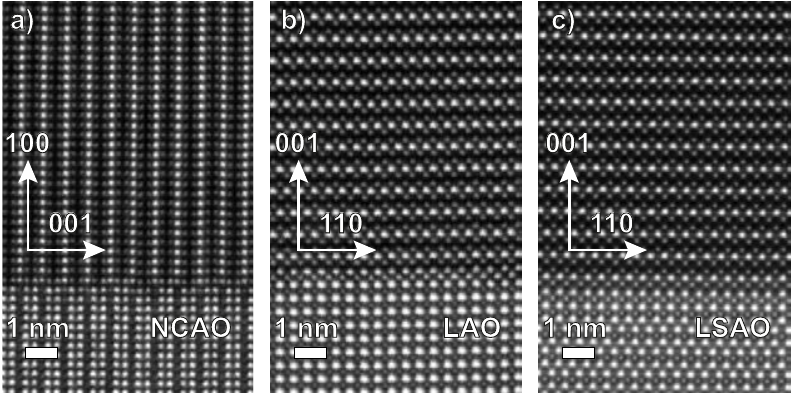}
	\caption{Cross-sectional HAADF images of CRO thin films on a) NCAO(110) b) LAO(100) and c)LSAO(001). Crystallographic directions refer to the orthorhombic unit cell of CRO.}
		\label{fig:TEM}
\end{figure}

\begin{table}
\centering
\begin{tabular}{l<{\hspace{1.5pc}}c<{\hspace{1pc}}c<{\hspace{1pc}}c<{\hspace{1pc}}l}
 &	a(\AA) & b(\AA) & c(\AA) & $\lvert \frac{a-b}{a+b} \rvert$ (\%) \\
\hline 
\hline \\ [.1pt]
S-CRO \SI{295}{\kelvin} \cite{Braden1998}	&	5.41		& 5.49		& 11.96	&	0.73 \\
S-CRO \SI{400}{\kelvin} \cite{Friedt2001}	&	5.36		& 5.35		& 12.26	&	0.09 \\
NCAO (110)				&	5.42 		& 5.30		& 12.11	&	1.12 \\
LAO	(100)				&	5.36 		& 5.36 		& 12.24	&	0.00 \\
LSAO (001)				&	5.31		& 5.31		& 12.37	&	0.00 \\
\hline 
\end{tabular}
\caption{Room-temperature lattice parameters of CRO thin films compared to literature values for bulk CRO below and above the MIT}
\label{tab:lt_const}
\end{table}

\begin{figure*}[t]
	\centering
		\includegraphics[width=\textwidth]{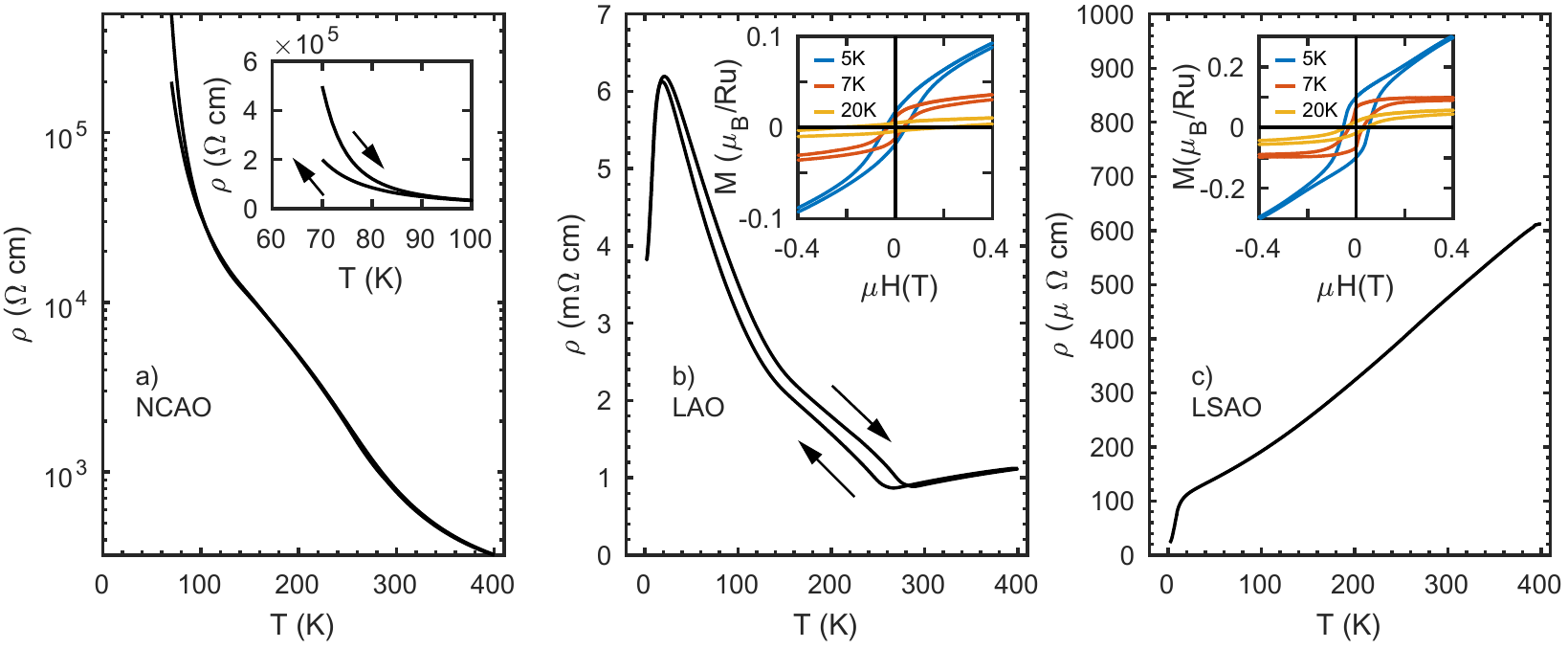}
	\caption{Temperature dependence of resistivity along the ab-planes of CRO films on a) NCAO b) LAO and c) LSAO. The inset of a) shows the resistivity hysteresis at low temperatures. The insets of b) and c) show the hysteresis curves of the magnetization in the low temperature FM phase measured along the in-plane [110] direction of CRO.}
		\label{fig:rt}
\end{figure*}

The structural characterization of the samples was conducted using  Cu-K$\mathrm{\alpha}$1 X-rays. Transport properties of the samples were studied using a Quantum Design PPMS setup via the standard four-point probe technique. Magnetometry data was acquired using a Quantum Design SQUID-VSM device. 

X-ray diffraction scans along the c-axis confirm the formation of a CRO phase on LSAO, LAO and NCAO substrates (Fig.\ref{fig:XRD_TTH}). Films similar to the samples shown here were also characterized at the MPI beamline at ANKA, Karlsruhe. \cite{Stierle2004a} All samples showed Bragg peaks linked to a doubling of the tetragonal unit cell indicating that the \ce{RuO_6} octahedra are tilted or rotated. Reflections found for CRO on LAO and LSAO can be indexed with a twinned structure based on the selection rules for Pbca, which is the space group for bulk CRO.\cite{Braden1998} CRO on NCAO (110) exhibits a single-domain structure, since the c-axis of the substrate naturally constitutes a preferred direction. The out-of-plane direction is compatible with the H0L selection rule of the Pbca space group allowing us to identify it as the a-axis. We note that, in addition to the expected peaks of the CRO phase, weak phi-dependent peaks were detected in scans along the 00L direction of CRO on LAO and LSAO  (see supplementary material for details). \pagebreak

Reciprocal Space Mapping (RSM) confirms that the films on LSAO and LAO are coherently strained to the substrate (Fig.\ref{fig:XRD_RSM}). The in-plane c-axis of the film on NCAO matches the c-axis of the substrate, whilst the in-plane b-axis as the other in-plane component is only partially strained. The fact that the system can only partly accommodate the compression along b and c is also visible in the RSM showing a typical triangular relaxation.\cite{Heinke1994} The c-axis lattice constants of the films on LAO and LSAO were determined by the specular out-of-plane XRD scans, while the in-plane constants are identical to the substrate due to the epitaxial relationship.\cite{Bueble1998,Kawamura2015} The a-axis lattice constant of the film on NCAO was derived from a $\theta-2\theta$ scan as well, while the in-plane lattice constants b and c were determined by refining a set of film peaks (Tab. \ref{tab:lt_const}).

The microstructure of the films was investigated using high-resolution transmission electron microscopy in High-Angle Annular Dark Field (HAADF) mode (Fig. \ref{fig:TEM}). Coherent growth is observed with only minor disorder in proximity to the interface (see supplementary material for low-magnification images and further details).

Single crystal CRO exhibits a large change in the lattice constants and volume at the MIT.\cite{Braden1998} The metallic high temperature phase (L-Pbca) is characterized by low orthorhombicity and a long c-axis ($c>\SI{12.2}{\angstrom}$) compared to the insulating low-temperature phase (Tab. \ref{tab:lt_const}). This correlation between electronic and structural properties is also found as a function of hydrostatic pressure.\cite{Steffens2005} In the following we will show that we can control the lattice properties of CRO via epitaxial strain and drive the material from an insulator to a metal by a systematic choice of substrates (Fig. \ref{fig:rt}). 

The result of our resistivity measurements of CRO on LAO match previously reported data on a \SI{170}{\nano\meter} film.\cite{Miao2012a} We find as well that films on LAO show a MIT at $T_{\textrm{MIT}}=\SI{290}{\kelvin}$ similar to single crystals, but the magnitude of the resistivity jump at the transition is much smaller than in bulk. \cite{Alexander1999} As conjectured in previous reports\cite{Miao2012a}, this might be related to the epitaxial relationship between film and substrate, which prevents CRO from completing the transition and keeps it in a metastable state as indicated by the broad hysteresis. The lattice constants of CRO on LAO in the metallic phase at room temperature are similar to the metallic phase in bulk CRO above the MIT (Tab. \ref{tab:lt_const}). Assuming that the in-plane lattice constants of CRO do not change significantly below the MIT due to the pinning of the substrate, the material can be considered to be \SIrange{1}{2}{\percent} compressively strained in the ab-plane relative to the insulating phase of the single crystal. The resistivity also shows a downturn at $T_{\textrm{C}}\approx \SI{20}{\kelvin}$ where the hysteresis concomitant with the MIT vanishes. Previous studies of CRO thin films on LAO and on pressurized single crystals attribute the cusp to a metallic FM phase.\cite{Taniguchi2013,Nakamura2002,Miao2012a} Except for the suppression of the MIT, its transport properties are similar to those of bulk CRO under \SIrange{0.5}{1}{\giga \pascal} hydrostatic pressure.\cite{Nakamura2007}

The structure of the film on NCAO is characterized by an orthorhombic distortion and the largest compression along the c-axis relative to the other films. Both distortions are likely to increase the tilt and the flattening of the \ce{RuO_6} octahedra, which have been linked to insulating CRO phases.\cite{Fang2001,Steffens2011} Indeed, the transport measurements show an insulating behavior from \SIrange{2}{400}{\kelvin}.  Relative to metallic bulk CRO at \SI{400}{\kelvin}, the material can be regarded as $\approx\SI{1}{\percent}$ compressively strained along the b- and c-axis. Despite showing no MIT up to \SI{400}{\kelvin}, we reproducibly measured a hysteresis starting at $\approx\SI{100}{\kelvin}$, which might be connected to a phase transition in the insulating state. Low temperature diffraction data are needed to clarify this issue.

The films grown on LSAO substrates show a metallic temperature dependence in the temperature range \SIrange{2}{400}{\kelvin}. Similar to CRO on LAO, a downturn at $T_{\textrm{C}}\approx\SI{20}{\kelvin}$ is found. The lattice constants of films on LSAO agree well with structural parameters known from metallic phases of bulk CRO.\cite{Steffens2005} Relative to the structure of insulating bulk CRO at \SI{295}{\kelvin}, the films on LSAO can be regarded to be $\SIrange{2}{3}{\percent}$ compressively strained in the ab-plane. The film on LSAO exhibits a residual resistivity of $\rho(\SI{2}{\kelvin})=\SI{30}{\micro\ohm\centi\meter}$, which is higher than the $\SI{3}{\micro\ohm\centi\meter}$ measured in pressurized single crystals.\cite{Nakamura2002} This difference can likely be attributed to defects caused by epitaxial strain or steps on the substrate surface and is similar to what has been observed when comparing \ce{SrRuO3} thin films with their single crystal counterpart.\cite{Thompson2016}
We verified the presence of ferromagnetism in the phase below the Curie temperature indicated by the transport measurements on a \SI{80}{\nano\meter} thick sample on LAO and the \SI{40}{\nano\metre} sample on LSAO via SQUID-VSM magnetometry. The insets in Fig. \ref{fig:rt} shows a typical ferromagnetic hysteresis in both samples below \SI{20}{\kelvin}. The data were corrected for the dominant linear diamagnetic response of the substrate at high fields. The saturated magnetic moment at \SI{5}{\kelvin} of CRO on LAO is $\approx \SI{0.1}{\mubohr}$ per Ru ion, whereas the saturated moment in the metallic film on LSAO appears to be approximately doubled at the same temperature. This is in line with observations on pressurized single crystals, where the moment increases from 0.1 to \SI{0.3}{\mubohr} in the pressure range \SIrange{0.3}{1.5}{\giga\pascal}.\cite{Nakamura2007,Nakamura2002}

\begin{figure}
	\centering
		\includegraphics[width=\columnwidth]{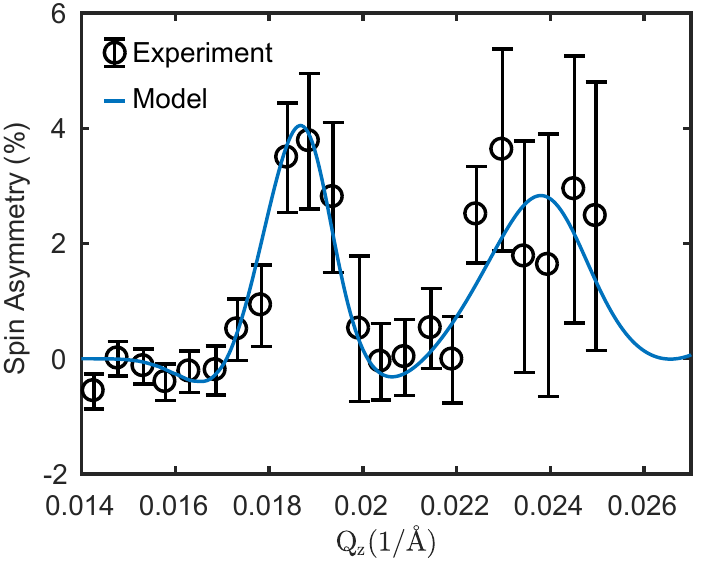}
	\caption{Spin asymmetry determined from Polarised Neutron Reflectometry. The sample was cooled in a field of \SI{4.5}{\kilo\oersted} along the [110] axis of CRO. The line is the result of a model calculation based on a homogeneous layer of CRO with a ferromagnetic moment of \SI{0.08}{\mubohr} per Ru.}
		\label{fig:PNR}
        \vspace*{-.2 cm}
\end{figure}

To further characterize the FM state, we performed polarized neutron reflectometry (PNR) experiments at the NREX reflectometer at FRM-II on the same \SI{80}{\nano\meter} thick CRO film on LAO.\cite{Khaydukov2015} PNR is a bulk-sensitive probe that yields depth-resolved magnetization profiles.\cite{Daillant2009} For this measurement, the sample was cooled to \SI{6}{\kelvin} in a field of \SI{4.5}{\kilo\oersted} applied along the [110] axis of CRO. The spin-resolved neutron reflectivities $R^{++}$ and $R^{--}$ were recorded using the neutron wavelength $\SI{4.28}{\angstrom}$. The resulting spin asymmetry $(R^{++}-R^{--})/(R^{++}+R^{--})$ was fitted with the SIMULREFLEC\footnote{SIMULREFLEC V1.7(2011) \url{http://www.llb.cea.fr/prism/programs/simulreflec/simulreflec.html}} software package, using structural parameters obtained by X-ray reflectivity (XRR) as starting values and literature values for the Neutron scattering lengths and X-ray scattering factors.\cite{Chantler1995, Sears1992} The XRR data were analyzed with the ReMagX software package.\cite{Macke2014c} A model based on the assumption that the complete film is magnetic yields an excellent description of the data (Fig.\ref{fig:PNR}). The resulting moment \SI{0.08}{\mubohr} per Ru is in good agreement with the magnetometry data. The PNR results thus confirm that the measured FM originates from the bulk of the film and not from impurities.

In summary, we have grown high-quality CRO films on NCAO, LAO and LSAO substrates and show that the epitaxial strain can profoundly modify the electronic ground state of the material from an insulator to a metal, spanning a range in resistivities over six orders of magnitude. Our study shows that increasing strain stabilizes a low temperature FM phase, analogous to the phase behavior of bulk CRO under pressure. Magnetometry and PNR measurements detect a low saturated moment in this phase, which is in line with the current picture that low-temperature FM phases in CRO are itinerant. The synthesis of CRO thin films on several substrates enables further studies of the magnetic and electronic structure under biaxial strain, and the large strain dependence of the electrical and magnetic properties may open up new perspectives for oxide electronics.

\section*{Supplementary}
See supplementary material for additional X-ray diffraction data, TEM images and the neutron reflectivity corresponding to the spin asymmetry shown in \linebreak Fig. \ref{fig:PNR}.

\begin{acknowledgments}
We thank E. Benckiser, J. Bertinshaw, M. Hepting, G. Kim, J. Porras and A. Seo for fruitful discussions, B. Lemke, U. Salzberger, B. Stuhlhofer and Y. Stuhlhofer for technical assistance. We acknowledge financial support by the European Research Council under Advanced Grant No. 669550 (Com4Com). This work is partially based upon experiments performed at the NREX instrument operated by the Max-Planck Society at the Heinz Maier-Leibnitz Zentrum (MLZ), Garching, Germany.
\end{acknowledgments}

\bibliography{bib_min}
\end{document}